\documentclass[prb,twocolumn,showpacs,preprintnumbers,amsmath,amssymb]{revtex4}
\usepackage{graphicx}% Include figure files
\usepackage{psfig}
\usepackage{dcolumn}% Align table columns on decimal point
\usepackage{bm}% bold math

\begin{document}

%\preprint{APS/123-QED}

\title{Magnetic and Transport Properties of La$_{0.7}$Sr$_{0.3}$Co$_{1-y}$Mn$_y$O$_3$}% Force line breaks with \\

\author{D.N.H. Nam}
 \altaffiliation[Also at ]{Physics Department, University of Cincinnati, Ohio}%Lines break automatically or can be forced with \\
 \email{namdao@physics.uc.edu}

\author{N.X. Phuc}%
\affiliation{%
Lab of Magnetism and Superconductivity, Institute of Materials
Science, NCST, Nghiado - Caugiay - Hanoi, Vietnam
}%

\author{L.V. Bau}%
 \altaffiliation[Also at ]{The \AA ngstr\"{o}m Laboratory, Uppsala University, Sweden.}

\author{N.V. Khiem}%

\affiliation{%
Department of Science and Technology, Hongduc University,
Thanhhoa, Vietnam
}%

\author{L.H. Son}
\affiliation{%
International Training Institute for Materials Science, 1A
Dai-Co-Viet, Hanoi, Vietnam
}%

\date{\today}% It is always \today, today,
             %  but any date may be explicitly specified

\begin{abstract}
While both La$_{0.7}$Sr$_{0.3}$MnO$_3$ and
La$_{0.7}$Sr$_{0.3}$CoO$_3$ are ferromagnets with metallic
conductivity below $T_\mathrm{c}$ (360K and 220K, respectively),
partial substitutions of Co by Mn in
La$_{0.7}$Sr$_{0.3}$Co$_{1-y}$Mn$_y$O$_3$ ($y < 0.1$) drastically
suppress the ferromagnetic long-range order pre-established by
Co-O-Co double exchange and tune the conductivity towards
insulating behavior. Since Mn-O-Mn double-exchange interactions
are avoidable at low Mn-substitution levels, the deterioration of
ferromagnetism and conductivity thus provides evidence for no
Mn-O-Co double exchange (but antiferromagnetic superexchange) in
the present system. At $y = 0.1$, the ferromagnetism is no longer
observed; the system becomes an insulating spin-glass with
$T_\mathrm{g}\approx 62$K as estimated from the ac-susceptibility
data using the conventional critical slowing-down scaling law.
With further substitution ($y\geq 0.3$), the ferromagnetism is
recovered ($T_\mathrm{c} = 165$K for $y = 0.3$ and 200K for $y =
0.5$) while the resistivity continues increasing and exhibits
insulating behavior. These results indicate that the substitution
is not simply a mixture of La$_{0.7}$Sr$_{0.3}$MnO$_3$ and
La$_{0.7}$Sr$_{0.3}$CoO$_3$, but produces a more complicated
scenario of spin states, interactions and disorder.
\end{abstract}

% PACS, the Physics and Astronomy
                             % Classification Scheme.
%\keywords{Suggested keywords}%Use showkeys class option if keyword
                              %display desired
\maketitle

\section{\label{sec:level1}Introduction}

In recent years, the hole-doped manganites $Ln_{1-x}A_x$MnO$_3$
($Ln$: Lanthanides, $A$: Alkaline earth elements) are one of the
most intensively studied topics, being strongly attractive to both
basic research and technology due to their rich physics and
potential prospects of application. Most of the undoped compounds,
$Ln$MnO$_3$, are insulators with an antiferromagnetic (AF) order
established by Mn$^{3+}$ ($t_{2g}^3e_g^1$, $S=2$) ions via AF
superexchange (SE) interactions below $T_\mathrm{N}$. The
substitution of $Ln^{3+}$ by $A^{2+}$ converts an adapted number
of Mn$^{3+}$ to Mn$^{4+}$ ($t_{2g}^3e_g^0$, $S=3/2$) giving rise
to ferromagnetic (FM) Mn$^{3+}$-O$^{2-}$-Mn$^{4+}$ double-exchange
(DE) interactions \cite{zener} where the transfer of the $e_g$
electron from Mn$^{3+}$ to Mn$^{4+}$ is favored by a parallel spin
configuration. In pertinent substitution ranges, where the DE
interactions dominate, some compounds could be metallic
ferromagnets and exhibit a colossal magneto-resistance (CMR)
phenomenon at $T_\mathrm{c}$. The DE mechanism is also usually
invoked to explain the magnetic and transport properties of the
hole-doped cobaltites, $Ln_{1-x}A_x$CoO$_3$, where the double
exchange is supposed to occur between Co$^{4+}$ and
Co$^{3+}$.\cite{itoh} Since the heart of the DE mechanism is the
mixed-valence state, it is then curious to know whether
double-exchange occurs between different transition-metal ions,
specially between Co and Mn ions. Furthermore, doping at $B$-sites
in manganites or cobaltites is also a good probe to study the DE
mechanism.

The magnetic and transport properties of
La$_{0.7}$Sr$_{0.3}$MnO$_3$ and La$_{0.7}$Sr$_{0.3}$CoO$_3$ have
been largely documented; both of them are double-exchange systems.
In this work, we study the magnetic and transport properties of
La$_{0.7}$Sr$_{0.3}$Co$_{1-y}$Mn$_y$O$_3$ where Co is partially
substituted by Mn. It is interesting that while the parent systems
are metallic ferromagnets ($T_\mathrm{c} = 360$K for
La$_{0.7}$Sr$_{0.3}$MnO$_3$, and 220K for
La$_{0.7}$Sr$_{0.3}$CoO$_3$), the mixed products become insulating
spin-glasses at certain substitution levels ($T_\mathrm{g}\approx
62$K for $y = 0.1$). The results also provide a clear evidence for
no DE interactions between Mn and Co ions.

\section{Experiment}

La$_{0.7}$Sr$_{0.3}$Co$_{1-x}$Mn$_{x}$O$_3$ samples with $y = 0$,
0.05, 0.1, 0.3, and 0.5 were prepared by conventional solid-state
reaction method using La$_2$O$_3$, SrCO$_3$, Co$_3$O$_4$, and
MnO$_2$ powders with purities of at least 99.9\% as raw materials.
The appropriate mixtures first underwent calcination at
1000$^\mathrm{o}$C for 24 hours. After being furnaced at
1300$^\mathrm{o}$C for another 24 hours, the products were then
sintered at 1350$^\mathrm{o}$C for 104 hours and slowly cooled to
room temperature at 3$^\mathrm{o}$C/min rate. The heat treatment
processes, always with very well pulverization and pelletization
in between, had all taken place in air. X-ray powder diffraction
measurements confirmed the single-phase property of all the
samples. The temperature dependence of ac magnetic
susceptibilities, $\chi'(T)$ and $\chi''(T)$, and resistivity,
$\rho(T)$, were measured using a Closed Cycle Helium Refrigerator.
The conventional four-probe method was used for $\rho(T)$
measurements except for $y = 0.3$ and 0.5, which have rather high
resistance, $\rho(T)$ data were read directly by a digital
multimeter. Temperature dependent field-cooled (FC) and
zero-field-cooled (ZFC) dc magnetizations, $M_{\mathrm{FC}}(T)$
and $M_{\mathrm{ZFC}}(T)$ respectively, were carried out by a
noncommercial vibrating-sample and a Quantum Design MPMS5
magnetometer.

\section {Results and Discussion}
Shown in Fig. 1 are the $M_{\mathrm{ZFC}}(T)$ and
$M_{\mathrm{FC}}(T)$ curves measured in an applied field of
$H_{\mathrm{ex}} = 100$G. For all the samples except $y = 0.1$,
which shows spin-glass behavior as shown in Fig. 4 and will be
discussed later, there exists a cusp in the $M_{\mathrm{ZFC}}(T)$
curves and a large separation of ZFC and FC magnetizations with
lowering temperature in the FM phase. This behavior could be
attributed to either (i) an appearance of frustration resulted
from the competition of SE and DE interactions or (ii) a gradually
freezing process of FM cluster moments caused by local anisotropy,
or perhaps both. In general, this behavior is always expected if
the measurements are carried out in an applied field smaller than
the saturation field that develops faster than the increase of the
saturation magnetization when the temperature is lowered. Quite
similar behaviors have been observed for
La$_{0.5}$Sr$_{0.5}$CoO$_3$ where the cusp of
$M_{\mathrm{ZFC}}(T)$ was explained by the competition between the
energies of anisotropy and external fields acting on the
ferromagnetic clusters existing in the system.\cite{nam} In fact,
La$_{0.7}$Sr$_{0.3}$CoO$_3$, and La$_{0.5}$Sr$_{0.5}$CoO$_3$ as
well, was classified as "cluster-glass" (CG) \cite{itoh} and we
have also experienced that the magnetic and transport properties
of the two compounds are qualitatively the same.

\begin{figure}
  % Requires \usepackage{graphicx}
  \vspace{0.1in}
  \includegraphics[width=2.8in]{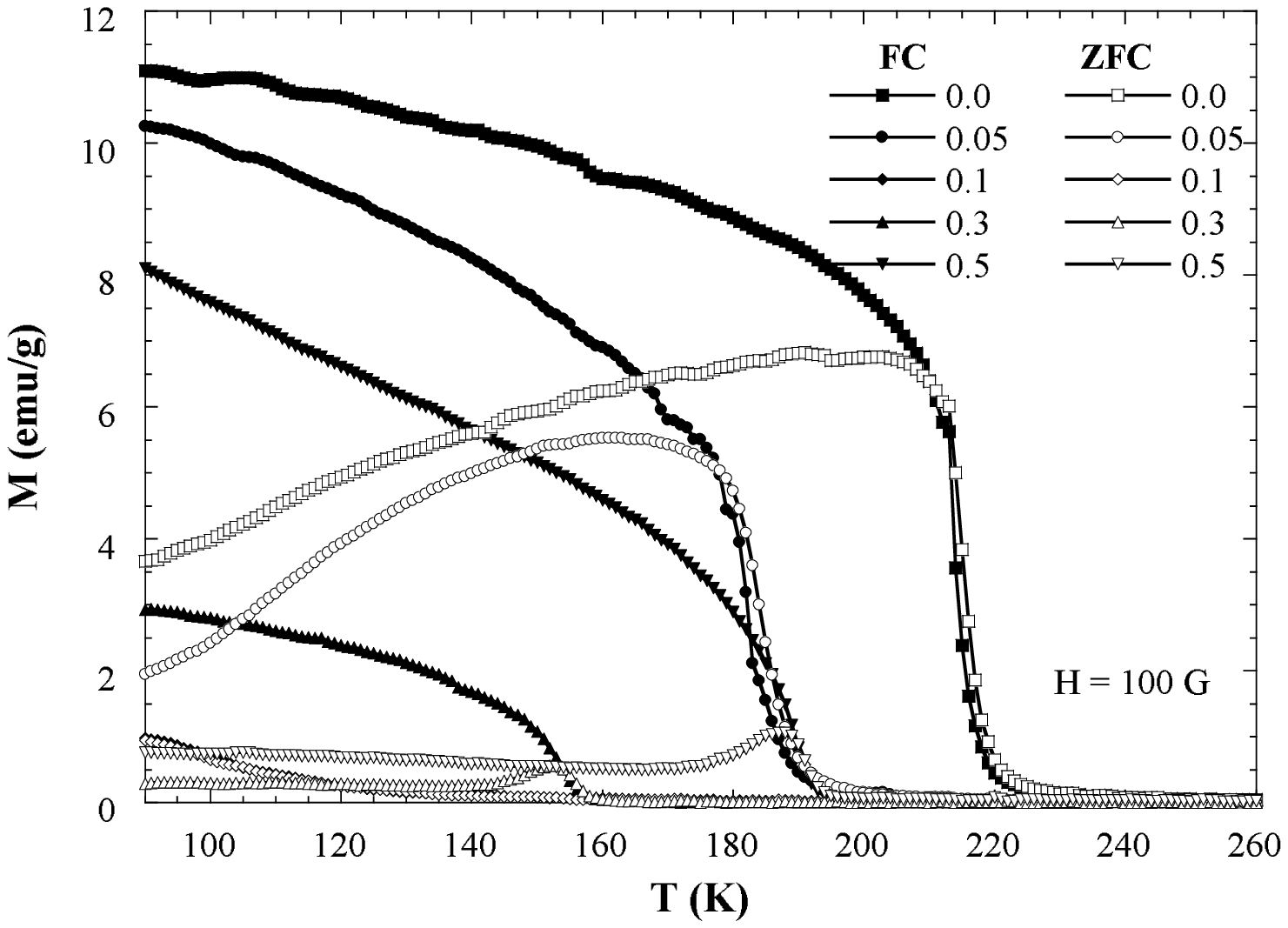}
  \caption{$M_{\mathrm{ZFC}}(T)$ (empty symbols) and $M_{\mathrm{FC}}(T)$
  (solid symbols) curves of La$_{0.7}$Sr$_{0.3}$Co$_{1-y}$Mn$_y$O$_3$
  samples with $y = 0$ ($\square$), 0.05 ($\circ$), 0.1 ($\diamond$), 0.3
  ($\vartriangle$), and 0.5 ($\triangledown$). $H = 100$ G.}\label{Figure1}
\end{figure}

\begin{figure}
  % Requires \usepackage{graphicx}
  \vspace{0.1in}
  \includegraphics[width=3.1in]{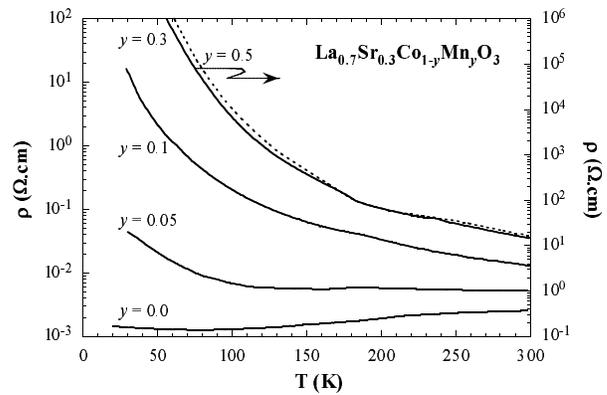}
  \caption{$\rho(T)$ curves of (bottom to top) the $y = 0.0$, 0.05,
  0.1 (left axis) and  $y = 0.3$, 0.5 (right axis) samples.}\label{Figure2}
\end{figure}

By fitting the susceptibility data, $\chi^{-1}(T)=H/M(T)$, at high
temperatures in the paramagnetic region to the Curie-Weiss law, we
find that the effective magnetic moment $\mu_{\mathrm{eff}}$ per
Co/Mn ion increases instantly with increasing $y$:
$\mu_{\mathrm{eff}} = 4.26$, 4.49, 4.89, and $5.09 \mathrm{\mu_B}$
for $y = 0$, 0.05, 0.1, and 0.3, respectively. It is interesting
that the variation of $\mu_{\mathrm{eff}}$ does not incorporate
with that of the phase transition temperature $T_\mathrm{c}$. We
attribute the increase of $\mu_{\mathrm{eff}}$ to stronger
cooperative Jahn-Teller distortions induced by the introduction of
Mn$^{3+}$ ions that lift the degeneracy of $t_{2g}$ and $e_g$
orbits of Co ions favoring them in higher-spin states.

\subsection{The undoped compound, $y = 0$}
The parent compound La$_{0.7}$Sr$_{0.3}$CoO$_3$ exhibits a
paramagnetic (PM) - ferromagnetic phase transition at
$T_\mathrm{c} = 220 \mathrm{K}$ in agreement with previously
reported data.\cite{itoh} For ferromagnetic manganites, the
conductivity is metallic below $T_\mathrm{c}$ but insulating above
$T_\mathrm{c}$, showing a metallic-insulating (MI) transition at
$T_\mathrm{c}$, in consistence with the DE mechanism. However, as
displayed in Fig. 2, the $\rho(T)$ curve of
La$_{0.7}$Sr$_{0.3}$CoO$_3$ merely changes the slope at
$T_\mathrm{c}$ and still exhibits metallic behavior in the
paramagnetic state. This feature is not expected for a typical DE
system but can be understood considering the fact that the
electronic configuration of the Co ions do not strictly obey
Hund's rule, an ingredient of the DE mechanism. That makes the
$e_g$ electron of the Co ions quite mobile in the sense that it
can hop between Co sites without strict requirement of a parallel
localized-$t_{2g}$-spin configuration or a high applied field. By
the way, this is one of the reasons for the magnetoresistance in
cobaltites usually small compared to the manganites.

Another remarkable feature in the $\rho(T)$ curve is an upward
"tail" at low temperature, showing an MI-like transition. In the
$\chi''(T)$ curve (not shown), we also observed a small "hump" at
$T_\mathrm{f}\sim 80$K signaling a transition to cluster-glass
state, a feature very similar to that observed for
La$_{0.5}$Sr$_{0.5}$CoO$_3$.\cite{nam} We assume that this
low-temperature resistance upturn is closely linked to the
transition to cluster-glass state, where the cluster moments are
randomly frozen,\cite{nam} since the metallicity according to DE
mechanism is not expected in such glassy states. This assumption
is in conformity with the fact that $T_\mathrm{f}$ is very close
to the temperature where the upturn begins. The formation of the
clusters is probably a consequence of an inhomogeneous
distribution of Sr$^{2+}$ ions.\cite{caciuffo} The bigger the
clusters, the higher the freezing temperature $T_\mathrm{f}$. In
order to check the influence of the inhomogeneity, we prepared a
series of samples with varying the sintering time $t_\mathrm{s}$.
The longer $t_\mathrm{s}$ would mean the better homogeneity and
therefore the finer clusters and lower $T_\mathrm{f}$. As
expected, the samples sintered by short times show a huge
low-temperature upturn with an increase of resistance by an order
while metallic behavior is still observed at high temperature. For
an example, Figure 3 displays the $\rho(T)$ and $\chi''(T)$ curves
of a $y = 0$ sample after being sintered for 5 h at
1350$^\mathrm{o}$C. Note from this figure that $T_\mathrm{f}$ and
the resistance upturn temperature are still close to each other
and both shift to higher temperatures ($\sim 150$K) as a result of
stronger inhomogeneity (i.e., larger cluster size) compared to
those of the 104h-sintered sample. The upturn is fading out in the
longer time sintered samples reducing to a minor tail (below $\sim
70$K) with $t_\mathrm{s} = 104$h as shown in Fig. 2.

However, because of the weak intra-atomic Hund's rule coupling of
the Co ions, the freezing of magnetic moments is unlikely the only
cause of the low-temperature resistance upturn although that
certainly mainly contributes. The same effect would be observed if
the sample contains insulating or semiconducting grain boundaries
where disorder and spin frustration are strong and a Coulomb gap
between small grains opens at low temperature.\cite{balcells}
Because the grain boundaries are improved and so is the grain size
by lasting $t_\mathrm{s}$, the two effects caused by grain
boundaries and cluster freezing are then mixed and
indistinguishable in the polycrystalline samples.

\subsection{The lightly doped compounds, $y = 0.05$ and $0.1$}

\begin{figure}
  % Requires \usepackage{graphicx}
  \vspace{0.1in}
  \includegraphics[width=3.0in]{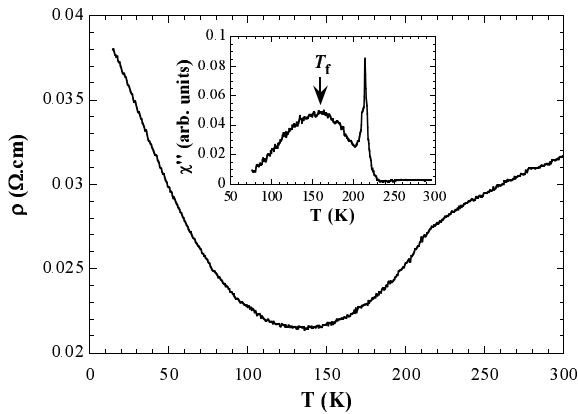}
  \caption{The $\rho(T)$ curve of a La$_{0.7}$Sr$_{0.3}$CoO$_3$ sample
  after being sintered at 1350$^\mathrm{o}$C for 5h. The inset shows
  $\chi''(T)$ measured at $f = 1$kHz. $T_\mathrm{f}$ marks the freezing temperature.}
  \label{Figure3}
\end{figure}

The influence of the substitution is so strong. $T_\mathrm{c}$
decreases from 220K for $y = 0$ to 192K for $y = 0.05$ and the
ferromagnetic state is no longer observed in the $y = 0.1$ sample.
Correspondingly, the resistivity increases strongly, and very
pronounced is the upturn at low temperatures that develops with
increasing Mn concentration and typical insulating behavior is
obviously seen for $y = 0.1$. Since the Mn ions are very dilute in
this substitution range, the substitution just creates new Mn-O-Co
bonds at the expense of the existing Co-O-Co bonds, implying that
there exists no Mn-O-Mn link. The severe deterioration of the
ferromagnetism and conductivity by such a low doping level as $y =
0.05$ provides a strong evidence for no DE interactions via
Mn-O-Co couplings. Moreover, these observations clearly indicate
that the Mn-O-Co interactions are superexchange antiferromagnetic
and much stronger than the DE Co-O-Co one. This conclusion is in
agreement with other previously reported experiments \cite{shi,
fan} where it was found that a substitution of Co for Mn also
severely suppresses the ferromagnetism and destroys the
metallicity of La$_{0.7}$Sr$_{0.3}$MnO$_3$. Increasing Mn
concentration results in an increase of the degree of frustration
and disorder and, at $y = 0.1$, the system becomes an insulating
spin-glass. Figure 4 presents the $M_{\mathrm{ZFC}}(T)$ and
$M_{\mathrm{FC}}(T)$ curves and the inset the $\chi'(T)$ curves of
the $y = 0.1$ sample measured at different frequencies $f =
0.037$, 0.37, 3.7, and 37 kHz. The results are very similar to
those of a conventional spin glass. With lowering temperature from
the paramagnetic phase, the spin relaxation of a spin-glass slows
down and the maximum relaxation time $\tau$ diverges at
$T_\mathrm{g}$ when the system enters the spin-glass state.
$\chi'$ attains a maximum at $T_\mathrm{f}$ that shifts towards
higher temperatures with higher frequencies. For the ac
susceptibilities, $T_\mathrm{f}$ defines the freezing temperature
where the characteristic time of the measurement, $\tau_\mathrm{m}
= 1/2\pi f$, becomes smaller than $\tau$ when the temperature is
lowered. An attempt at fitting the $T_\mathrm{f}(f)$ data
extracted from Fig. 4 to conventional slowing-down scaling law,
$\tau/\tau_\mathrm{o}=[(T_\mathrm{f}-T_\mathrm{g})/T_\mathrm{g}]^{-z\nu}$,
gives $\tau_\mathrm{o} = 1.1\times 10^{-13}$ s, $z\nu = 8.2$, and
$T_\mathrm{g} = 61.8$ K, implying that the
La$_{0.7}$Sr$_{0.3}$Co$_{0.9}$Mn$_{0.1}$O$_3$ compound is a
conventional 3D spin-glass.

\begin{figure}
  % Requires \usepackage{graphicx}
  \vspace{0.1in}
  \includegraphics[width=3.0in]{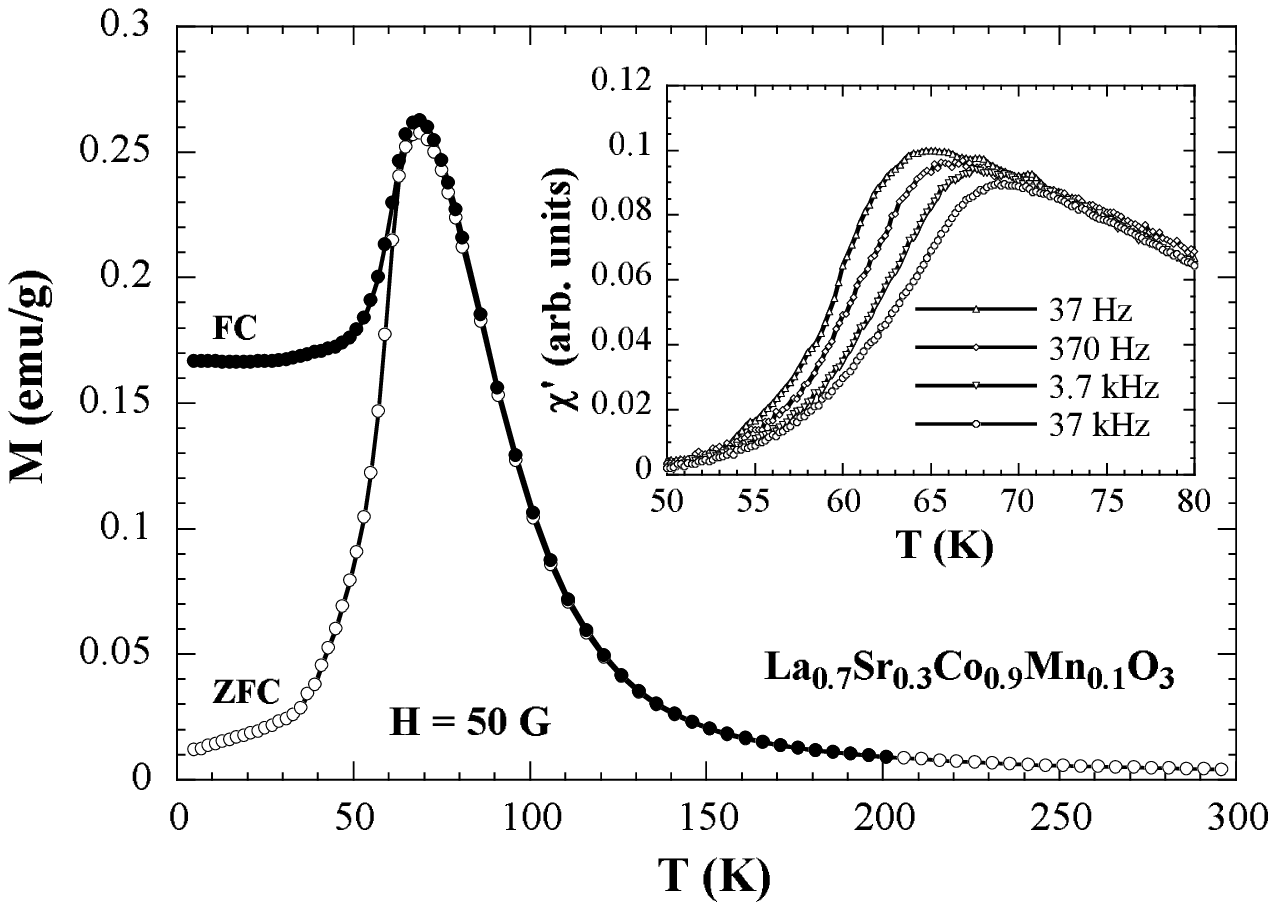}
  \caption{Temperature dependence of $M_{\mathrm{ZFC}}$ and $M_{FC}$
  of the $y=0.1$ sample in applied field $H = 50$G. Inset: $\chi''(T)$
  measured at $f = 37$Hz ($\triangle$), 370Hz ($\diamond$), 3.7kHz ($\nabla$), and 37kHz ($\circ$).}
  \label{Figure4}
\end{figure}

There is also a remarkable feature in the evolution towards
insulating behavior of the system, as seen in Fig. 2, where the
low-temperature resistance upturn dominantly develops with
increasing $y$. The metallicity totally disappears once the system
becomes a spin-glass in consistence with the DE mechanism. This
indicates that the upturn in the Mn-doped compounds is closely
related to the weakening of ferromagnetic correlation and the
appearance of frustration and disorder, consolidating the above
assumption that it has a tight relationship with the freezing of
cluster moments.

The Mn ions are believed to always exist in high-spin states
whereas, because of the comparable magnitude of the crystal field
splitting and the Hund's coupling energy, the Co ions can exist in
different spin states. The effective magnetic moment value of
4.26$\mathrm{\mu_B}$ found for $y = 0$ does not allow us to
extract qualitative information about the spin states of Co ions.
However, qualitatively, $\mu_{\mathrm{eff}} = 4.26\mathrm{\mu_B}$
implies that a part of Co ions, either of trivalence or
tetravalence, or both, must exist in intermediate- and/or low-spin
states. When increasing $y$ from 0 to 0.05, the ferromagnetism is
suppresses, however, $\mu_{\mathrm{eff}}$ increases by an amount
of 0.23$\mathrm{\mu_B}$, that is almost equal to the possible
maximum contribution of 0.25 $\mathrm{\mu_B}$ from the doped Mn
ions in the case that all of them are Mn$^{3+}$. This sounds as if
(i) all Mn ions have replaced to only the low-spin state
Co$^{\mathrm{III}}$ sites or (ii) an amount of Co ions must have
transferred to higher-spin states. Moreover, for the case of $y =
0.1$, the gain in $\mu_{\mathrm{eff}}$ is 0.63$\mathrm{\mu_B}$
while the possible maximum contribution from Mn ions (i.e.,
assuming they are all Mn$^{3+}$) is only 0.49$\mathrm{\mu_B}$.
This later case unambiguously indicates that, in this doping
region, there are certainly significant amounts of Co ions
converted to higher spin-states as a consequence of the
substitution. Since Mn$^{3+}$ are strong Jahn-Teller ions, it is
possible that an introduction of Mn$^{3+}$ into the system by the
substitution induces stronger Jahn-Teller distortions lifting the
degeneracy of $t_{2g}$ and $e_g$ orbits of Co ions, that favors a
conversion of the Co ions to higher-spin states.

\subsection {The heavily doped compounds, $y = 0.3$ and $0.5$}

With further substitution the ferromagnetism reappears at $y =
0.3$ with $T_\mathrm{c} = 165$K, that increases to 200K for $y =
0.5$. When the Mn ions are dense, the probability for them to
interact to each other is high. Assuming that the Mn ions exist in
a mixed-valence state of Mn$^{3+}$ and Mn$^{4+}$, the recovery of
the ferromagnetism is most probably due to the appearance of
ferromagnetic Mn$^{3+}$-O-Mn$^{4+}$ DE interactions. In contrast,
although the ferromagnetism is rapidly recovered, the resistivity
continues increasing and exhibiting insulating behavior. This
implies that the FM DE interactions are still strongly competed by
AF SE interactions, and not strong enough to establish continuous
pathways for the system to exhibit metallic behavior. A plausible
explanation is that, in this doping range, the magnetic as well as
electronic ordering is not uniform; the ferromagnetism signal
comes from (metallic) ferromagnetic regions embedded in an
insulating (non-ferromagnetic) matrix. For higher Mn
concentrations, it was found in earlier experiments that, in this
substitution range up to $y = 1$, $T_\mathrm{c}$ monotonously
increases, and the metallic conductivity is gradually recovered
when $y$ being close to 1.\cite{shi, fan}

$\mu_{\mathrm{eff}}$ increases strongly with increasing $y$ from 0
to 0.1 and then slows down in the region $0.1 \leq y\leq 0.3$. For
the case of the lightly doped compounds, an over gain of
$\mu_{\mathrm{eff}}$ has told us that a number of Co ions must be
converted to their higher-spin states. Nevertheless, that is not
the case for the heavily doped compounds where the gain in
$\mu_{\mathrm{eff}}$ is moderate with respect to the substitution
levels. Fortunately, the high values of $\mu_{\mathrm{eff}}$
suggest that all of the Co ions are in high-spin states. For the
case of $y = 0.3$, assuming that all Co and Mn ions are in
high-spin states and the ratios Co$^{3+}$/Co$^{4+}$ and
Mn$^{3+}$/Mn$^{4+}$ are kept at 7/3, we obtain $\mu_{\mathrm{eff}}
= 5.02\mathrm{\mu_B}$, a value very close to that experimentally
determined of 5.09$\mathrm{\mu_B}$. These facts lead to a
suggestion that most of the Co ions are in high-spin states and no
further spin-state conversion occurs when $y$ goes above 0.3.

\section{Conclusion}

The magnetic and transport properties of
La$_{0.7}$Sr$_{0.3}$Co$_{1-y}$Mn$_y$O$_3$ have been systematically
studied and discussed in detail. The Mn substitution induces a
conversion of Co ions to higher-spin states. The magnetic and
transport properties of the system are tightly correlated. The
obtained results provide a clear evidence for no double-exchange
interaction between Mn and Co ions but strong antiferromagnetic
superexchange ones. As a result, the competition of AF and FM
interactions introduced by the substitution lead the system to
become an insulating spin-glass at $y = 0.1$. For the samples with
high Mn concentrations, $y = 0.3$ and 0.5, the magnetic as well as
electronic phase is not uniform; (metallic) ferromagnetic clusters
of Mn$^{3+}$-O$^{2-}$-Mn$^{4+}$ bonds are formed on a
(non-ferromagnetic) insulating background. We have also evidenced
that a transition to glassy state could in principle destroy the
ferromagnetic state metallicity of a double exchange ferromagnet.

\begin{acknowledgments}
We wish to thank Dr. P.V. Phuc for x-ray diffraction measurements
and analyses. Financial supports from SIDA/Sarec Project are
highly acknowledged.
\end{acknowledgments}

\end{document}